\documentclass[twocolumn,aps,prl,preprintnumbers,groupedaddress,tightenlines,nofootinbib,floatfix,10pt,reprint,superscriptaddress]{revtex4-2}

\usepackage{amssymb,amsmath,hyperref,xcolor}
\hypersetup{
    colorlinks=true,
    linkcolor=blue,
    citecolor=blue,
    urlcolor=blue,
}

\usepackage{bm}
\usepackage{color,xcolor}
\usepackage{soul}
\usepackage{placeins}
\usepackage{graphicx,multirow,tabularx}
\usepackage{longtable}

\newcommand{\Nsite}{N_{\rm site}}
\newcommand{\Nc}{N_{\rm c}}
\newcommand{\Nf}{N_{\rm f}}
\newcommand{\Nt}{N_t}
\newcommand{\Ns}{N_s}

\newcommand{\Tr}{\mathrm{Tr}}

\begin{document}

\preprint{YITP-26-83, J-PARC-TH-0337}

\title{Higher-order hopping-parameter expansion by human-AI collaboration}

% ===== Authors (order follows Downloads/main.tex: Kitazawa, Wada) =====
\author{Masakiyo Kitazawa}
\email[]{kitazawa@yukawa.kyoto-u.ac.jp}
\affiliation{
  Yukawa Institute for Theoretical Physics, Kyoto University, Kyoto, 606-8502, Japan}
\affiliation{
  J-PARC Branch, KEK Theory Center, Institute of Particle and Nuclear
  Studies, KEK, Tokai, Ibaraki 319-1106, Japan}

\author{Tatsuya Wada}
\email[]{tatsuya.wada@yukawa.kyoto-u.ac.jp}
\affiliation{
  Yukawa Institute for Theoretical Physics, Kyoto University, Kyoto, 606-8502, Japan}

\date{\today}

\begin{abstract}
We develop efficient algorithms for evaluating higher-order terms in the hopping-parameter expansion of $\Tr\ln M$ on $SU(\Nc)$ gauge configurations. The resulting algorithms, which exploit a trie data structure for the computation of high-order terms, evaluate the $\kappa^8$, $\kappa^{10}$, and $\kappa^{12}$ terms at computational costs of approximately $20$, $460$, and $8900$ times that of a single staple evaluation, respectively. 
The correctness of the algorithms is verified by comparison with a computationally expensive but reliable reference calculation. We emphasize that collaboration between human researchers and AI coding agents was essential to the development of these algorithms.
\end{abstract}

\maketitle

\section{Introduction}

Lattice quantum chromodynamics (QCD) plays a unique role in providing first-principles calculations for quantitative studies of strongly interacting systems
~\cite{BMW:2008jgk,PACS-CS:2009sof,Edwards:2011jj,Aoki:2012tk,Fodor:2012gf,Borsanyi:2013bia,HotQCD:2014kol,RBC:2015gro,Chang:2018uxx,Yang:2018nqn,Borsanyi:2020mff,Aoyama:2024cko,FLAG:2024oxs,Davies:2025pmx,Borsanyi:2025ttb,Boccaletti:2024guq}.
% ~\cite{FLAG:2024oxs,Aoyama:2024cko,Ding:2015ona,Guenther:2020jwe,Philipsen:2019rjq, Aoki:2006we,Bazavov:2011nk, HotQCD:2014kol,HotQCD:2019xnw,Borsanyi:2020fev,Borsanyi:2021sxv, deForcrand:2002hgr, Allton:2005gk}. 
As lattice simulations continue to advance toward the continuum limit, larger physical volumes, and higher precision, improving computational efficiency has become increasingly important. Recent advances in generative artificial intelligence (AI) have opened new opportunities for scientific computing, and its application to lattice QCD simulations has attracted growing interest~\cite{Matsumoto:2019jia,Boyda:2020hsi, Abbott:2022zhs, Favoni:2020reg, Nicoli:2020njz, Wang:2023exq, Boyda:2022nmh, Albergo:2021vyo, Cranmer:2023xbe,Tomiya:2025quf,Yasunaga:2026smj,Misumi:2026zbe}.

Besides numerical simulations, lattice field theory provides analytic approaches based on perturbative expansions, such as the strong-coupling expansion and the hopping-parameter expansion (HPE)~\cite{Wilson:1974sk,Munster:1980vk,Drouffe:1980dp,Munster:1981es,Creutz_2023,Rothe:1992nt}. These methods are useful not only for gaining insight into qualitative features of the theory, but also as practical ingredients in exact Monte Carlo calculations, improving statistical precision~\cite{Gulpers:2013uca,Zhou:2018,Hasenbusch:2018iuw,Baral:2019mkt,Giusti:2019kff,Djukanovic:2023beb}, and for studying systems under extreme conditions, such as the heavy-quark region of QCD~\cite{Philipsen:2014rpa,Cuteri:2020yke,Philipsen:2021qji,Kiyohara:2021smr,Ashikawa:2024njc}. A characteristic feature of these expansions is that higher-order terms are represented by geometric objects on the lattice, such as trajectories and surfaces. Their classification becomes increasingly challenging at higher orders because of the rapid combinatorial growth of structures. In the present Letter, we focus on the HPE and illustrate that AI can play a useful role in developing efficient algorithms for such computations.

In the HPE, fermionic contributions for Wilson fermions are expressed as a power series in the hopping parameter $\kappa$. The coefficient of $\kappa^n$ is represented by length-$n$ closed trajectories on the four-dimensional lattice. At the leading ($\kappa^4$) and next-to-leading ($\kappa^6$) orders, the trajectories without temporal windings are classified into only one and three distinct types, respectively (see Fig.~\ref{fig:trajectories}). Their evaluation has long been well understood~\cite{Rothe:1992nt} and widely used in numerical studies~\cite{Gulpers:2013uca,Zhou:2018,Hasenbusch:2018iuw,Baral:2019mkt,Giusti:2019kff,Djukanovic:2023beb,Kiyohara:2021smr,Ashikawa:2024njc}. At higher orders, however, the number of trajectories grows rapidly, making their evaluation increasingly difficult. To the best of our knowledge, no explicit computation of higher-order terms on gauge configurations beyond order $\kappa^6$ has been reported in the literature.

In the present Letter, we show that higher-order terms in the HPE can be evaluated at practical computational cost by developing explicit algorithms to compute them on $SU(\Nc)$ gauge configurations for arbitrary $\Nc$. The key idea is to exploit a trie data structure~\cite{Fredkin1960} to organize trajectories so that matrix multiplications (MMs) shared by multiple trajectories are evaluated only once. This method substantially reduces the computational cost at high orders.
In our current implementation, the number of MMs required for the $\kappa^{10}$ and $\kappa^{12}$ terms is approximately $460$ and $8900$ times that of a single staple evaluation, respectively, which remain modest compared with many numerical procedures in lattice QCD. We also implement the algorithm for $\Nc=3$, which is publicly available~\cite{WadaHPECode}, and confirm that the wall-clock time is well described by the number of MMs. In contrast, the conventional shape-based algorithm remains more efficient for lower-order terms.

A distinctive feature of the present work is that the algorithm for the $\kappa^8$ term was developed by human researchers, whereas the principal algorithmic innovation for higher orders, the trie-based algorithm, was discovered by an AI coding agent. This finding by AI, however, was built upon the trajectory classification and shape-based algorithms developed by humans during the earlier stages. Another crucial ingredient was a human-developed, computationally expensive but reliable reference implementation, which enabled rigorous validation of candidate algorithms. We devote a later section to documenting the development process of this project, as we expect that it will be beneficial for the future development of computational science.

\section{Hopping-parameter expansion}
\label{sec:setup}

One of the standard choices for relativistic lattice fermions is the Wilson fermion, whose fermion matrix in four-dimensional spacetime is given by 
\begin{align}
    &M_\kappa(x,y) = \delta_{x,y} - \kappa\, B(x,y),
    \label{eq:fermion_matrix}\\
    &B(x,y) 
    \notag \\
    &= (1-\gamma_4)U_{x,4} e^{\hat\mu_q}\,\delta_{y,x+\hat4}
        +(1+\gamma_4)U^\dagger_{y,4} e^{-\hat\mu_q}\,\delta_{y,x-\hat4}
        \notag \\
        &+\sum_{i=1}^3\Big[
        (1-\gamma_i)U_{x,i}\,\delta_{y,x+\hat i}
        +(1+\gamma_i)U^\dagger_{y,i}\,\delta_{y,x-\hat i}
    \Big],
    \label{eq:hopping_term}
\end{align}
with the gamma matrices $\gamma_{1,2,3,4}$ and the dimensionless chemical potential $\hat\mu_q$. The link variables $U_{x,\mu}$ are $SU(\Nc)$ matrices with the number of colors $\Nc$ and the hopping parameter $\kappa$ is related to the bare-quark mass $m_0$ as $\kappa= 1/(2m_0a+8)$ with the lattice spacing $a$. For QCD with $\Nf$ flavors, integrating out the fermionic degrees of freedom yields the partition function
\begin{align}
    Z = \int\mathcal{D}U\,
    \exp \Big[ -S_g + \sum_f\,\ln\textrm{Det} M_{\kappa_f} \Big],
    \label{eq:Z}
\end{align}
with the gauge action $S_g$ and $\textrm{Det}$ being the determinant over all indices of $M$. The summation over $f$ runs over all flavors, where $\kappa_f$ denotes the hopping parameter of the $f$th flavor.

The fermionic contribution in Eq.~\eqref{eq:Z}, $\ln\textrm{Det} M_\kappa=\Tr\ln M_\kappa$, is expanded in powers of $\kappa$ as
\begin{align}
  \Tr\ln M_\kappa = \Nsite\sum_{n} C_n\,\kappa^n,
  &&
  C_n = -\frac{1}{\Nsite\,n}\,\Tr\!\left[B^{\,n}\right],
  \label{eq:HPE}
\end{align}
where $\Nsite = \Ns^3\times\Nt$ is the spacetime volume of the lattice with spatial and temporal extents $\Ns$ and $\Nt$, and $\Tr$ denotes the trace over the same space as $\textrm{Det}$. Equation~\eqref{eq:HPE} is known as the HPE of $\Tr\ln M$. Since the limit $\kappa\to0$ corresponds to the heavy-quark limit $m_0\to\infty$, the HPE is regarded as a heavy quark-mass expansion. Since $B$ takes nonzero values only between nearest-neighbor lattice sites, each coefficient $C_n$ is represented by the sum of the products of link variables along connected closed lattice trajectories of length $n$~\cite{Rothe:1992nt,Gattringer:2010zz}. For free fermions with $U_{x,\mu}=1$ on all link variables, the radius of convergence of Eq.~\eqref{eq:HPE} is $\kappa=1/8$~\cite{Wakabayashi:2021eye}, which suggests that the HPE may remain convergent even on gauge-field backgrounds up to the chiral limit.

For $\hat\mu_q\ne0$, $C_n$ can be decomposed into trajectories having different temporal winding numbers $\ell$ as 
\begin{align}
  C_n = W(n) + \sum_{\ell=1}^{\infty} \big( L_\ell(N_t,n) e^{\ell N_t \hat\mu_q} + L_{-\ell}(N_t,n) e^{-\ell N_t\hat\mu_q} \big) \,,
  \label{eq:loopex}
\end{align}
where $W(n)$ collects the Wilson loops without temporal windings, while $L_\ell(N_t,n)$ contains the Polyakov-loop-type (PLT) loops that wind in the temporal direction $\ell$ times; $W(n)$ vanish for $n<4$ and odd $n$, while $L_\ell(N_t,n)$ vanish for $n < \ell N_t$ by definition. In what follows, we refer to $W(4)$ and $L_\ell(N_t,N_t)$ as the leading-order (LO) terms of the Wilson and PLT loops, $W(6)$ and $L_\ell(N_t,N_t+2)$ as the next-to-leading-order (NLO) terms, etc. Note that the order in $\kappa$ is different for $W(n)$ and $L_\ell(N_t,n)$ in this definition except for $N_t=4$.

\begin{figure}
    \centering
    \includegraphics[width=.98\linewidth]{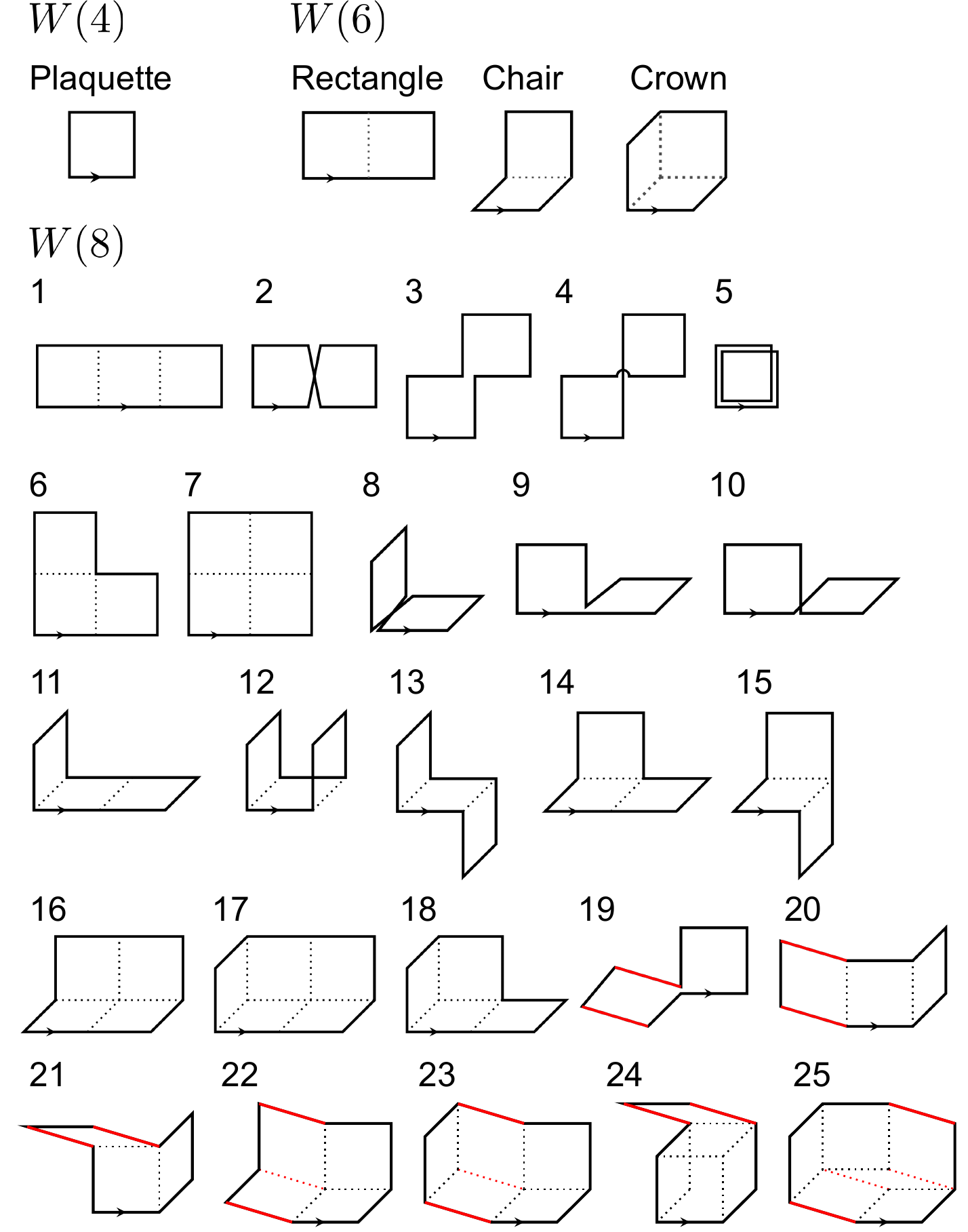}
    \caption{Trajectories included in $W(4)$, $W(6)$, and $W(8)$. Red lines denote links extending in the fourth spatial direction.}

    \label{fig:trajectories}
\end{figure}

To calculate each term in Eq.~\eqref{eq:loopex}, it is convenient to classify the contributions from individual trajectories by their shapes as
\begin{align}
    W(n) &= - 2\Nc \sum_j \frac{M_j D_j}{S_j} {\rm Re}\, \mathcal{W}_j(n) \,,
    \label{eq:W} \\
    L_\ell (N_t,n) &= - (-1)^\ell \Nc \sum_j \frac{M_j D_j}{S_j} \mathcal{L}_j (N_t,n,\ell)\,,
    \label{eq:L}
\end{align}
where $j$ represents the different shapes and the summation runs over all shapes in each term. For instance, $W(4)$ contains only the ``plaquette'', while $W(6)$ consists of three types, ``rectangle'', ``chair'', and ``crown'', shown in Fig.~\ref{fig:trajectories}. The trajectories in $W(8)$ are classified into the $25$ shapes shown in the figure, where the trajectories having ``appendices''~\cite{Rothe:1992nt} are excluded since they do not contribute to the HPE.
$M_j$ denotes the number of distinct trajectories belonging to class $j$ per lattice site under cubic rotations/reflections (CuRR) forming the hypercubic group $H(4)$ and $H(3)$ for $W(n)$ and $L_\ell(N_t,n)$, respectively, and $D_j$ is the coefficient from the trace in the Dirac space $D_j = {\rm tr_D} [ (1-\gamma^\mu)(1-\gamma^\nu) \cdots]$, where ${\rm tr_D}$ denotes the trace over the Dirac indices.
$\mathcal{W}_j(n)$ and $\mathcal{L}_j(N_t,n,\ell)$ represent the products of link variables defined, for example, by $\mathcal{W}_j(n) = \sum_{\textrm{ all trajectories in}~j}^{M_j N_{\rm site}} \textrm{tr}_{\rm c}[U_{x_1,\mu}U_{x_2,\nu}\cdots ]/(\Nc\Nsite M_j)$, where $\textrm{tr}_{\rm c}$ denotes the trace over the color indices, and the normalization is imposed so that $\mathcal{W}_j(n)$ and $\mathcal{L}_j(N_t,n,\ell)$ become unity for free fermions.
Finally, $S_j$ denotes the symmetry factor, defined as the number of cyclic permutations (CyP) that leave a trajectory invariant.

Note that $W(n)$ contains each trajectory together with its reverse. Since they are complex conjugates of each other, $W(n)$ is real. The factor of $2$ in Eq.~\eqref{eq:W} arises from this degeneracy, while each pair is counted only once in $M_j$. On the other hand, $L_\ell(N_t,n)$ is complex in general with $L_\ell(N_t,n)=L_{-\ell}(N_t,n)^*$. The factor $(-1)^\ell$ in Eq.~\eqref{eq:L} comes from the anti-periodic temporal boundary condition for fermions. More detailed definitions and derivations of Eqs.~\eqref{eq:W} and~\eqref{eq:L} can be found in Refs.~\cite{Rothe:1992nt,Kiyohara:2021smr,Ashikawa:2024njc,Wakabayashi:2021eye}.

\section{Classification of trajectories}

As a first step toward the evaluation of higher-order terms, we classify the trajectories in $W(n)$ and $L_\ell(N_t,n)$ by their shapes.

A trajectory of length $n$ on a four-dimensional lattice is represented by a sequence of lattice directions, $\{\pm1,\pm2,\pm3,\pm4\}$, of length $n$. 
For $W(n)$, the same number of $\pm\mu$ must appear to form a closed path. For $L_\ell(N_t,n)$, the number of forward-temporal steps, $+4$, must exceed that of $-4$ by $\ell N_t$. 
Trajectories that become identical under CyP, reversal, and CuRR are classified into the same shape class.

In our project, we first developed an algorithm for this classification manually and then improved it with assistance from an AI agent (\texttt{Claude Opus 4.8}). The resulting program classifies trajectories at N$^4$LO in less than one minute on a laptop computer. Once the classification is completed, it is straightforward to calculate $M_j$, $D_j$, and $S_j$ for each shape. We also note that the orbit-counting theorem, known as Burnside’s lemma~\cite{Burnside:1897}, provides a validation of this procedure.

Table~\ref{tab:trajectory} summarizes the number of distinct shapes in each term, $N_{\rm shape}$, and the total number of different trajectories per lattice site, $N_{\rm traj}=\sum_{j=1}^{N_{\rm shape}}M_j/S_j$, for $W(n)$ and $L_\ell(N_t,n)$ for $N_t=6$, where shapes with $D_j=0$ are excluded from their definitions. Both $N_{\rm shape}$ and $N_{\rm traj}$ grow rapidly with $n$.
The complete lists of shapes and corresponding values of $M_j$, $D_j$, $S_j$ are provided in the supplementary material up to N$^2$LO and Ref.~\cite{WadaHPECode}.

\begin{table}[t]
\caption{Number of distinct shapes, $N_{\rm shape}$, and the total number of different trajectories per lattice site, $N_{\rm traj}=\sum_{j=1}^{N_{\rm shape}}M_j/S_j$, for $W(n)$ and $L_\ell(N_t,n)$ with $N_t=6$. In the latter, the values for the sum over positive $\ell$ are shown. The shapes with $D_j=0$ are excluded from these numbers.
}
\label{tab:trajectory}
\begin{ruledtabular}
\begin{tabular}{lrr}
& $N_{\rm shape}$ & $N_{\rm traj}$ \\
\hline
$W(4)$ & 1 & 6 \\
$W(6)$ & 3 & 76 \\
$W(8)$ & 24 & $1.71\cdot10^3$ \\%1,713 \\
$W(10)$ & 189 & $3.80\cdot10^4$\\%38,040 \\
$W(12)$ & 3701 & $1.03\cdot10^6$ \\%1,031,788 \\
%\hline
%$L(4,4)$ & 1 & 0.25 \\
%$L(4,6)$ & 2 & 9 \\
%$L(4,8)$ & 19 & 259.625 \\
%$L(4,10)$ & 338 & 9,114 \\
%$L(4,12)$ & 10,104 & 363,524.5833 \\
\hline
$L(6,6)$ & 1 & 1/6 \\
$L(6,8)$ & 3 & 15 \\
$L(6,10)$& 56 & 870 \\
$L(6,12)$& 1477 & $4.19\cdot10^4$\\%41879.083333333336
$L(6,14)$& 56395 & $2.06\cdot10^6$ \\
\end{tabular}
\end{ruledtabular}
\end{table}

The values of $W(n)$ and $L_\ell(N_t,n)$ for free fermions have been calculated in Ref.~\cite{Wakabayashi:2021eye} up to extremely high orders; see also App.~A of Ref.~\cite{Ashikawa:2024njc}. These results must be reproduced by substituting $\mathcal{W}_j(n)=1$ and $\mathcal{L}_j(N_t,n,\ell)=1$ into Eqs.~\eqref{eq:W} and~\eqref{eq:L}. This property serves as another validation of the classification.

\section{Reference program}

As a reference for subsequent algorithmic development, we implemented a reliable but computationally expensive program. This program uses the shape-classification list obtained in the previous section, which contains, for each $j$, its representative trajectory and the corresponding value of $D_j$. It computes the trace of matrix products for all trajectories in class $j$ generated from the representative trajectory by CyP and CuRR at every lattice site. $W(n)$ and $L_\ell(N_t,n)$ are then obtained by substituting these results into Eqs.~\eqref{eq:W} and~\eqref{eq:L}. Because this implementation requires access to at most $n+1$ neighboring lattice sites along a single direction for N$^n$LO, this implementation is not well suited for MPI parallelization at large $n$.

For later convenience, let us estimate the computational cost of this algorithm. The dominant computational cost arises from MMs. For a given trajectory, computing the trace of the product of link variables along it requires $n-2$ MMs. The resulting cost for $W(n)$ and $L_\ell(N_t,n)$ is therefore approximately $(n-2)N_{\rm traj}$~MM/site. Taking the cost of a staple evaluation~\cite{Rothe:1992nt,Gattringer:2010zz}, $1~{\rm staple}=48~{\rm MM/site}$ as the unit, Table~\ref{tab:trajectory} shows that the computational costs of $W(8)$ and $W(10)$ are approximately 
$2.1\cdot10^2$ and $6.3\cdot10^3~{\rm staples}$,
respectively. These costs may be acceptable for various applications, while the latter is already demanding.

\section{Efficient algorithms}

To evaluate higher-order terms at practical computational cost, we developed an efficient implementation that enables the evaluation up to N$^4$LO ($W(12)$ and $L_\ell(N_t,N_t+8)$) on a given gauge configuration. The algorithm assumes the unitarity of the link variables, but is applicable to arbitrary $\Nc$.

Our implementation beyond N$^2$LO is based on a trie data structure~\cite{Fredkin1960}, which exploits the fact that many lattice trajectories share identical partial paths. The algorithm first enumerates all trajectories obtained by applying CuRR to every representative trajectory in the shape-classification list. Each trajectory is then divided into two halves of equal length, and they are independently organized into trie structures. Matrix products associated with common trie nodes are evaluated only once and reused in subsequent multiplications of subtrajectories. After constructing the matrix products for the two halves, the value for the closed loop is obtained by taking the trace of their product. The full implementation is available in Ref.~\cite{WadaHPECode}.

This strategy substantially reduces the number of MMs compared with the reference implementation. The resulting computational costs are approximately $28$, $460$ and $8900$~staples, for $W(8)$, $W(10)$, and $W(12)$, respectively. The costs for $L_\ell(N_t,N_t+n)$ show similar scaling. In our implementation for $\Nc=3$~\cite{WadaHPECode}, we found that the actual execution time is almost proportional to the number of MMs. Although the precise computational cost depends on the choice of the splitting point of trajectories, this dependence is weak in our tests. The required memory for this algorithm is independent of the lattice volume.

On the other hand, for N$^2$LO we found that the trie-based algorithm is less efficient than a shape-based strategy. The latter optimizes the evaluation of each shape by performing partial summations over subtrajectories to reduce the number of MMs. It also exploits precomputed ``single-staples'' of the form $U_{x,\mu}U_{x+\hat\mu,\nu}U^\dagger_{x+\hat\nu,\mu}$ and untraced plaquettes. The best implementation of this strategy evaluates $W(8)$ at a cost of about $20$ staples~\cite{WadaHPECode}.

Besides the computational cost, MPI communication is another important issue for large-scale simulations. A na\"ive implementation of the trie algorithm requires access to at most $n+1$ neighboring lattice sites along a given direction at N$^n$LO. Choosing the splitting point appropriately, the required depth can be reduced to $\lceil (n+1)/2 \rceil$. For N$^2$LO and N$^3$LO, this corresponds to two halo layers in each direction. With the use of the single-staples, the number of halo layers can be reduced further. No MPI communication is necessary during the evaluation of $W(n)$ and $L_\ell(N_t,n)$, except for the communication of single-staples.

\section{Development process and role of AI}

As AI systems become increasingly important in scientific research, we believe it is valuable to document the development process of the present study, focusing on the complementary roles of human researchers and AI coding agents.

The project was originally initiated with the goal of constructing a program to evaluate the N$^2$LO terms, $W(8)$ and $L_\ell(N_t,N_t+4)$, by human effort alone. In this stage, we first developed a program for the trajectory classification, followed by optimized routines based on the shape-based algorithm. The resulting implementation reduced the computational cost to approximately $20$ and $8$ staples for $W(8)$ and $L_\ell(4,8)$, respectively. Part of this development history is available in Ref.~\cite{WadaHPECode}.

At this point, we introduced the AI coding agents (\texttt{Claude Code/Claude Opus 4.8} and \texttt{ChatGPT Codex/ChatGPT GPT-5.5}). They were tasked with improving the existing N$^2$LO implementation and extending it to N$^3$LO, where the shape-classification list, reference program, and the development history of human-written codes were provided~\cite{WadaHPECode}. The AI agents then proposed the trie-based algorithm and produced an implementation for N$^3$LO, whose correctness was verified using the reference program. 

Thus, one of the key advances of this study,  the trie-based algorithm, was introduced by AI. Interestingly, instead of refining the shape-based strategy as we had anticipated, the AI proposed a conceptually different algorithm, which, in retrospect, appears to be a natural formulation of the problem.

We, however, believe that this success also relies on the preparation by human researchers. In particular, the shape-classification list contributed to the discovery of the trie-based algorithm by AI, and the human-written reference code enabled rigorous validation of AI-generated code. To examine this point, we tested whether AI could develop an efficient program directly from the definitions of the HPE. In our experience, this approach did not produce meaningful results. Moreover, we have not succeeded in teaching AI the optimization for MPI environments. At present, the procedure to specify the splitting point for reducing communication layers is imposed manually by human researchers. 

These experiences suggest that AI is a powerful source of algorithmic ideas, but successful research still depends on human researchers to formulate the problem, prepare reliable validation tools, and refine AI-generated results.

\section{Discussion}

The algorithms developed in the present study are directly applicable to various numerical calculations in lattice QCD. In addition to applications to heavy-quark QCD~\cite{Kiyohara:2021smr,Ashikawa:2024njc}, higher-order HPE coefficients can be used to reduce statistical uncertainties in stochastic estimators for $\Tr\ln M$ and the fermion bilinear $\langle\bar\psi\Gamma\psi\rangle  =\Tr[\Gamma M^{-1}] =\sum_n\Tr[\Gamma B^n]\kappa^n$~\cite{Gulpers:2013uca,Zhou:2018,Hasenbusch:2018iuw,Baral:2019mkt,Giusti:2019kff,Djukanovic:2023beb}. For $\Gamma=1$, this expansion shares the same coefficients $C_n$ as Eq.~\eqref{eq:HPE}. Other choices of $\Gamma$ can also be accommodated simply by replacing the coefficients $D_j$ in Eqs.~\eqref{eq:W} and~\eqref{eq:L}. Efficient evaluation of higher-order HPE coefficients will also be useful in reducing statistical uncertainties of two-point correlation functions, including both connected and disconnected contributions.

The algorithms presented here leave room for further improvement. One promising direction is a hybrid approach combining the trie-based and shape-based algorithms, in which each trajectory is evaluated by a more efficient method. Another natural extension is the evaluation of still higher-order terms, which is feasible in principle. It is also interesting to explore other perturbative expansions, such as the strong-coupling expansion~\cite{Munster:1980vk,Creutz_2023}, with AI assistance.
We leave these directions for future work.

In the present Letter, we have shown that higher-order terms in the HPE on $SU(\Nc)$ gauge configurations can be evaluated at practical computational costs up to N$^4$LO by developing efficient algorithms based on a trie data structure. The resulting algorithms enable the exact evaluation of higher-order HPE coefficients and open new opportunities for practical applications, such as the reduction of statistical uncertainties in stochastic estimators and the numerical investigation of heavy-quark QCD. Beyond this technical advance, the present work documents how the algorithmic advance emerged through human-AI collaboration. We hope that this experience will provide a useful example for future research in computational physics.

\begin{acknowledgments}

We thank Kazuyuki Kanaya, Takahiro M. Doi, Yoshimasa Hidaka and Keito Shimizu for useful discussions and encouragement. This work was supported in part by JSPS KAKENHI Grant Numbers JP22K03619, JP24K07049, ISHIZUE 2025 of Kyoto University, and the Center for Gravitational Physics and Quantum Information (CGPQI) at Yukawa Institute for Theoretical Physics. T.~W. is supported by Grant-in-Aid for JSPS Fellows
No.26KJ1395. We acknowledge the assistance of the AI coding agents \texttt{Claude Code} (\texttt{Claude Opus 4.8}) and \texttt{ChatGPT Codex} (\texttt{ChatGPT GPT-5.5}) in algorithm development, code implementation, and manuscript preparation.

\end{acknowledgments}

\bibliography{bibliography}

\begin{thebibliography}{50}%
\makeatletter
\providecommand \@ifxundefined [1]{%
 \@ifx{#1\undefined}
}%
\providecommand \@ifnum [1]{%
 \ifnum #1\expandafter \@firstoftwo
 \else \expandafter \@secondoftwo
 \fi
}%
\providecommand \@ifx [1]{%
 \ifx #1\expandafter \@firstoftwo
 \else \expandafter \@secondoftwo
 \fi
}%
\providecommand \natexlab [1]{#1}%
\providecommand \enquote  [1]{``#1''}%
\providecommand \bibnamefont  [1]{#1}%
\providecommand \bibfnamefont [1]{#1}%
\providecommand \citenamefont [1]{#1}%
\providecommand \href@noop [0]{\@secondoftwo}%
\providecommand \href [0]{\begingroup \@sanitize@url \@href}%
\providecommand \@href[1]{\@@startlink{#1}\@@href}%
\providecommand \@@href[1]{\endgroup#1\@@endlink}%
\providecommand \@sanitize@url [0]{\catcode `\\12\catcode `\$12\catcode `\&12\catcode `\#12\catcode `\^12\catcode `\_12\catcode `\%12\relax}%
\providecommand \@@startlink[1]{}%
\providecommand \@@endlink[0]{}%
\providecommand \url  [0]{\begingroup\@sanitize@url \@url }%
\providecommand \@url [1]{\endgroup\@href {#1}{\urlprefix }}%
\providecommand \urlprefix  [0]{URL }%
\providecommand \Eprint [0]{\href }%
\providecommand \doibase [0]{https://doi.org/}%
\providecommand \selectlanguage [0]{\@gobble}%
\providecommand \bibinfo  [0]{\@secondoftwo}%
\providecommand \bibfield  [0]{\@secondoftwo}%
\providecommand \translation [1]{[#1]}%
\providecommand \BibitemOpen [0]{}%
\providecommand \bibitemStop [0]{}%
\providecommand \bibitemNoStop [0]{.\EOS\space}%
\providecommand \EOS [0]{\spacefactor3000\relax}%
\providecommand \BibitemShut  [1]{\csname bibitem#1\endcsname}%
\let\auto@bib@innerbib\@empty
%</preamble>
\bibitem [{\citenamefont {Durr}\ \emph {et~al.}(2008)\citenamefont {Durr} \emph {et~al.}}]{BMW:2008jgk}%
  \BibitemOpen
  \bibfield  {author} {\bibinfo {author} {\bibfnamefont {S.}~\bibnamefont {Durr}} \emph {et~al.} (\bibinfo {collaboration} {BMW}),\ }\bibfield  {title} {\bibinfo {title} {{Ab-Initio Determination of Light Hadron Masses}},\ }\href {https://doi.org/10.1126/science.1163233} {\bibfield  {journal} {\bibinfo  {journal} {Science}\ }\textbf {\bibinfo {volume} {322}},\ \bibinfo {pages} {1224} (\bibinfo {year} {2008})},\ \Eprint {https://arxiv.org/abs/0906.3599} {arXiv:0906.3599 [hep-lat]} \BibitemShut {NoStop}%
\bibitem [{\citenamefont {Aoki}\ \emph {et~al.}(2010)\citenamefont {Aoki} \emph {et~al.}}]{PACS-CS:2009sof}%
  \BibitemOpen
  \bibfield  {author} {\bibinfo {author} {\bibfnamefont {S.}~\bibnamefont {Aoki}} \emph {et~al.} (\bibinfo {collaboration} {PACS-CS}),\ }\bibfield  {title} {\bibinfo {title} {{Physical Point Simulation in 2+1 Flavor Lattice QCD}},\ }\href {https://doi.org/10.1103/PhysRevD.81.074503} {\bibfield  {journal} {\bibinfo  {journal} {Phys. Rev. D}\ }\textbf {\bibinfo {volume} {81}},\ \bibinfo {pages} {074503} (\bibinfo {year} {2010})},\ \Eprint {https://arxiv.org/abs/0911.2561} {arXiv:0911.2561 [hep-lat]} \BibitemShut {NoStop}%
\bibitem [{\citenamefont {Edwards}\ \emph {et~al.}(2011)\citenamefont {Edwards}, \citenamefont {Dudek}, \citenamefont {Richards},\ and\ \citenamefont {Wallace}}]{Edwards:2011jj}%
  \BibitemOpen
  \bibfield  {author} {\bibinfo {author} {\bibfnamefont {R.~G.}\ \bibnamefont {Edwards}}, \bibinfo {author} {\bibfnamefont {J.~J.}\ \bibnamefont {Dudek}}, \bibinfo {author} {\bibfnamefont {D.~G.}\ \bibnamefont {Richards}},\ and\ \bibinfo {author} {\bibfnamefont {S.~J.}\ \bibnamefont {Wallace}},\ }\bibfield  {title} {\bibinfo {title} {{Excited state baryon spectroscopy from lattice QCD}},\ }\href {https://doi.org/10.1103/PhysRevD.84.074508} {\bibfield  {journal} {\bibinfo  {journal} {Phys. Rev. D}\ }\textbf {\bibinfo {volume} {84}},\ \bibinfo {pages} {074508} (\bibinfo {year} {2011})},\ \Eprint {https://arxiv.org/abs/1104.5152} {arXiv:1104.5152 [hep-ph]} \BibitemShut {NoStop}%
\bibitem [{\citenamefont {Aoki}\ \emph {et~al.}(2012)\citenamefont {Aoki}, \citenamefont {Doi}, \citenamefont {Hatsuda}, \citenamefont {Ikeda}, \citenamefont {Inoue}, \citenamefont {Ishii}, \citenamefont {Murano}, \citenamefont {Nemura},\ and\ \citenamefont {Sasaki}}]{Aoki:2012tk}%
  \BibitemOpen
  \bibfield  {author} {\bibinfo {author} {\bibfnamefont {S.}~\bibnamefont {Aoki}}, \bibinfo {author} {\bibfnamefont {T.}~\bibnamefont {Doi}}, \bibinfo {author} {\bibfnamefont {T.}~\bibnamefont {Hatsuda}}, \bibinfo {author} {\bibfnamefont {Y.}~\bibnamefont {Ikeda}}, \bibinfo {author} {\bibfnamefont {T.}~\bibnamefont {Inoue}}, \bibinfo {author} {\bibfnamefont {N.}~\bibnamefont {Ishii}}, \bibinfo {author} {\bibfnamefont {K.}~\bibnamefont {Murano}}, \bibinfo {author} {\bibfnamefont {H.}~\bibnamefont {Nemura}},\ and\ \bibinfo {author} {\bibfnamefont {K.}~\bibnamefont {Sasaki}} (\bibinfo {collaboration} {HAL QCD}),\ }\bibfield  {title} {\bibinfo {title} {{Lattice QCD approach to Nuclear Physics}},\ }\href {https://doi.org/10.1093/ptep/pts010} {\bibfield  {journal} {\bibinfo  {journal} {PTEP}\ }\textbf {\bibinfo {volume} {2012}},\ \bibinfo {pages} {01A105} (\bibinfo {year} {2012})},\ \Eprint {https://arxiv.org/abs/1206.5088} {arXiv:1206.5088 [hep-lat]} \BibitemShut {NoStop}%
\bibitem [{\citenamefont {Fodor}\ and\ \citenamefont {Hoelbling}(2012)}]{Fodor:2012gf}%
  \BibitemOpen
  \bibfield  {author} {\bibinfo {author} {\bibfnamefont {Z.}~\bibnamefont {Fodor}}\ and\ \bibinfo {author} {\bibfnamefont {C.}~\bibnamefont {Hoelbling}},\ }\bibfield  {title} {\bibinfo {title} {{Light Hadron Masses from Lattice QCD}},\ }\href {https://doi.org/10.1103/RevModPhys.84.449} {\bibfield  {journal} {\bibinfo  {journal} {Rev. Mod. Phys.}\ }\textbf {\bibinfo {volume} {84}},\ \bibinfo {pages} {449} (\bibinfo {year} {2012})},\ \Eprint {https://arxiv.org/abs/1203.4789} {arXiv:1203.4789 [hep-lat]} \BibitemShut {NoStop}%
\bibitem [{\citenamefont {Borsanyi}\ \emph {et~al.}(2014)\citenamefont {Borsanyi}, \citenamefont {Fodor}, \citenamefont {Hoelbling}, \citenamefont {Katz}, \citenamefont {Krieg},\ and\ \citenamefont {Szabo}}]{Borsanyi:2013bia}%
  \BibitemOpen
  \bibfield  {author} {\bibinfo {author} {\bibfnamefont {S.}~\bibnamefont {Borsanyi}}, \bibinfo {author} {\bibfnamefont {Z.}~\bibnamefont {Fodor}}, \bibinfo {author} {\bibfnamefont {C.}~\bibnamefont {Hoelbling}}, \bibinfo {author} {\bibfnamefont {S.~D.}\ \bibnamefont {Katz}}, \bibinfo {author} {\bibfnamefont {S.}~\bibnamefont {Krieg}},\ and\ \bibinfo {author} {\bibfnamefont {K.~K.}\ \bibnamefont {Szabo}},\ }\bibfield  {title} {\bibinfo {title} {{Full result for the QCD equation of state with 2+1 flavors}},\ }\href {https://doi.org/10.1016/j.physletb.2014.01.007} {\bibfield  {journal} {\bibinfo  {journal} {Phys. Lett. B}\ }\textbf {\bibinfo {volume} {730}},\ \bibinfo {pages} {99} (\bibinfo {year} {2014})},\ \Eprint {https://arxiv.org/abs/1309.5258} {arXiv:1309.5258 [hep-lat]} \BibitemShut {NoStop}%
\bibitem [{\citenamefont {Bazavov}\ \emph {et~al.}(2014)\citenamefont {Bazavov} \emph {et~al.}}]{HotQCD:2014kol}%
  \BibitemOpen
  \bibfield  {author} {\bibinfo {author} {\bibfnamefont {A.}~\bibnamefont {Bazavov}} \emph {et~al.} (\bibinfo {collaboration} {HotQCD}),\ }\bibfield  {title} {\bibinfo {title} {{Equation of state in ( 2+1 )-flavor QCD}},\ }\href {https://doi.org/10.1103/PhysRevD.90.094503} {\bibfield  {journal} {\bibinfo  {journal} {Phys. Rev. D}\ }\textbf {\bibinfo {volume} {90}},\ \bibinfo {pages} {094503} (\bibinfo {year} {2014})},\ \Eprint {https://arxiv.org/abs/1407.6387} {arXiv:1407.6387 [hep-lat]} \BibitemShut {NoStop}%
\bibitem [{\citenamefont {Bai}\ \emph {et~al.}(2015)\citenamefont {Bai} \emph {et~al.}}]{RBC:2015gro}%
  \BibitemOpen
  \bibfield  {author} {\bibinfo {author} {\bibfnamefont {Z.}~\bibnamefont {Bai}} \emph {et~al.} (\bibinfo {collaboration} {RBC, UKQCD}),\ }\bibfield  {title} {\bibinfo {title} {{Standard Model Prediction for Direct CP Violation in K{\textrightarrow}{\ensuremath{\pi}}{\ensuremath{\pi}} Decay}},\ }\href {https://doi.org/10.1103/PhysRevLett.115.212001} {\bibfield  {journal} {\bibinfo  {journal} {Phys. Rev. Lett.}\ }\textbf {\bibinfo {volume} {115}},\ \bibinfo {pages} {212001} (\bibinfo {year} {2015})},\ \Eprint {https://arxiv.org/abs/1505.07863} {arXiv:1505.07863 [hep-lat]} \BibitemShut {NoStop}%
\bibitem [{\citenamefont {Chang}\ \emph {et~al.}(2018)\citenamefont {Chang} \emph {et~al.}}]{Chang:2018uxx}%
  \BibitemOpen
  \bibfield  {author} {\bibinfo {author} {\bibfnamefont {C.~C.}\ \bibnamefont {Chang}} \emph {et~al.},\ }\bibfield  {title} {\bibinfo {title} {{A per-cent-level determination of the nucleon axial coupling from quantum chromodynamics}},\ }\href {https://doi.org/10.1038/s41586-018-0161-8} {\bibfield  {journal} {\bibinfo  {journal} {Nature}\ }\textbf {\bibinfo {volume} {558}},\ \bibinfo {pages} {91} (\bibinfo {year} {2018})},\ \Eprint {https://arxiv.org/abs/1805.12130} {arXiv:1805.12130 [hep-lat]} \BibitemShut {NoStop}%
\bibitem [{\citenamefont {Yang}\ \emph {et~al.}(2018)\citenamefont {Yang}, \citenamefont {Liang}, \citenamefont {Bi}, \citenamefont {Chen}, \citenamefont {Draper}, \citenamefont {Liu},\ and\ \citenamefont {Liu}}]{Yang:2018nqn}%
  \BibitemOpen
  \bibfield  {author} {\bibinfo {author} {\bibfnamefont {Y.-B.}\ \bibnamefont {Yang}}, \bibinfo {author} {\bibfnamefont {J.}~\bibnamefont {Liang}}, \bibinfo {author} {\bibfnamefont {Y.-J.}\ \bibnamefont {Bi}}, \bibinfo {author} {\bibfnamefont {Y.}~\bibnamefont {Chen}}, \bibinfo {author} {\bibfnamefont {T.}~\bibnamefont {Draper}}, \bibinfo {author} {\bibfnamefont {K.-F.}\ \bibnamefont {Liu}},\ and\ \bibinfo {author} {\bibfnamefont {Z.}~\bibnamefont {Liu}},\ }\bibfield  {title} {\bibinfo {title} {{Proton Mass Decomposition from the QCD Energy Momentum Tensor}},\ }\href {https://doi.org/10.1103/PhysRevLett.121.212001} {\bibfield  {journal} {\bibinfo  {journal} {Phys. Rev. Lett.}\ }\textbf {\bibinfo {volume} {121}},\ \bibinfo {pages} {212001} (\bibinfo {year} {2018})},\ \Eprint {https://arxiv.org/abs/1808.08677} {arXiv:1808.08677 [hep-lat]} \BibitemShut {NoStop}%
\bibitem [{\citenamefont {Borsanyi}\ \emph {et~al.}(2021)\citenamefont {Borsanyi} \emph {et~al.}}]{Borsanyi:2020mff}%
  \BibitemOpen
  \bibfield  {author} {\bibinfo {author} {\bibfnamefont {S.}~\bibnamefont {Borsanyi}} \emph {et~al.},\ }\bibfield  {title} {\bibinfo {title} {{Leading hadronic contribution to the muon magnetic moment from lattice QCD}},\ }\href {https://doi.org/10.1038/s41586-021-03418-1} {\bibfield  {journal} {\bibinfo  {journal} {Nature}\ }\textbf {\bibinfo {volume} {593}},\ \bibinfo {pages} {51} (\bibinfo {year} {2021})},\ \Eprint {https://arxiv.org/abs/2002.12347} {arXiv:2002.12347 [hep-lat]} \BibitemShut {NoStop}%
\bibitem [{\citenamefont {Aoyama}\ \emph {et~al.}(2024)\citenamefont {Aoyama}, \citenamefont {Doi}, \citenamefont {Doi}, \citenamefont {Itou}, \citenamefont {Lyu}, \citenamefont {Murakami},\ and\ \citenamefont {Sugiura}}]{Aoyama:2024cko}%
  \BibitemOpen
  \bibfield  {author} {\bibinfo {author} {\bibfnamefont {T.}~\bibnamefont {Aoyama}}, \bibinfo {author} {\bibfnamefont {T.~M.}\ \bibnamefont {Doi}}, \bibinfo {author} {\bibfnamefont {T.}~\bibnamefont {Doi}}, \bibinfo {author} {\bibfnamefont {E.}~\bibnamefont {Itou}}, \bibinfo {author} {\bibfnamefont {Y.}~\bibnamefont {Lyu}}, \bibinfo {author} {\bibfnamefont {K.}~\bibnamefont {Murakami}},\ and\ \bibinfo {author} {\bibfnamefont {T.}~\bibnamefont {Sugiura}} (\bibinfo {collaboration} {HAL QCD}),\ }\bibfield  {title} {\bibinfo {title} {{Scale setting and hadronic properties in the light quark sector with (2+1)-flavor Wilson fermions at the physical point}},\ }\href {https://doi.org/10.1103/PhysRevD.110.094502} {\bibfield  {journal} {\bibinfo  {journal} {Phys. Rev. D}\ }\textbf {\bibinfo {volume} {110}},\ \bibinfo {pages} {094502} (\bibinfo {year} {2024})},\ \Eprint {https://arxiv.org/abs/2406.16665} {arXiv:2406.16665 [hep-lat]} \BibitemShut {NoStop}%
\bibitem [{\citenamefont {Aoki}\ \emph {et~al.}(2026)\citenamefont {Aoki} \emph {et~al.}}]{FLAG:2024oxs}%
  \BibitemOpen
  \bibfield  {author} {\bibinfo {author} {\bibfnamefont {Y.}~\bibnamefont {Aoki}} \emph {et~al.} (\bibinfo {collaboration} {Flavour Lattice Averaging Group (FLAG)}),\ }\bibfield  {title} {\bibinfo {title} {{FLAG review 2024}},\ }\href {https://doi.org/10.1103/nfzp-p5dn} {\bibfield  {journal} {\bibinfo  {journal} {Phys. Rev. D}\ }\textbf {\bibinfo {volume} {113}},\ \bibinfo {pages} {014508} (\bibinfo {year} {2026})},\ \Eprint {https://arxiv.org/abs/2411.04268} {arXiv:2411.04268 [hep-lat]} \BibitemShut {NoStop}%
\bibitem [{\citenamefont {Davies}(2025)}]{Davies:2025pmx}%
  \BibitemOpen
  \bibfield  {author} {\bibinfo {author} {\bibfnamefont {C.}~\bibnamefont {Davies}},\ }\bibfield  {title} {\bibinfo {title} {{Muon g-2}},\ }\href {https://doi.org/10.22323/1.466.0019} {\bibfield  {journal} {\bibinfo  {journal} {PoS}\ }\textbf {\bibinfo {volume} {LATTICE2024}},\ \bibinfo {pages} {019} (\bibinfo {year} {2025})},\ \Eprint {https://arxiv.org/abs/2503.03364} {arXiv:2503.03364 [hep-lat]} \BibitemShut {NoStop}%
\bibitem [{\citenamefont {Borsanyi}\ and\ \citenamefont {Parotto}(2025)}]{Borsanyi:2025ttb}%
  \BibitemOpen
  \bibfield  {author} {\bibinfo {author} {\bibfnamefont {S.}~\bibnamefont {Borsanyi}}\ and\ \bibinfo {author} {\bibfnamefont {P.}~\bibnamefont {Parotto}},\ }\href@noop {} {\bibinfo {title} {{The QCD phase diagram}}} (\bibinfo {year} {2025}),\ \Eprint {https://arxiv.org/abs/2512.08843} {arXiv:2512.08843 [nucl-th]} \BibitemShut {NoStop}%
\bibitem [{\citenamefont {Boccaletti}\ \emph {et~al.}(2026)\citenamefont {Boccaletti} \emph {et~al.}}]{Boccaletti:2024guq}%
  \BibitemOpen
  \bibfield  {author} {\bibinfo {author} {\bibfnamefont {A.}~\bibnamefont {Boccaletti}} \emph {et~al.},\ }\bibfield  {title} {\bibinfo {title} {{Hybrid calculation of hadronic vacuum polarization in muon g {\ensuremath{-}} 2 to 0.48{\%}}},\ }\href {https://doi.org/10.1038/s41586-026-10449-z} {\bibfield  {journal} {\bibinfo  {journal} {Nature}\ }\textbf {\bibinfo {volume} {653}},\ \bibinfo {pages} {373} (\bibinfo {year} {2026})},\ \Eprint {https://arxiv.org/abs/2407.10913} {arXiv:2407.10913 [hep-lat]} \BibitemShut {NoStop}%
\bibitem [{\citenamefont {Matsumoto}\ \emph {et~al.}(2021)\citenamefont {Matsumoto}, \citenamefont {Kitazawa},\ and\ \citenamefont {Kohno}}]{Matsumoto:2019jia}%
  \BibitemOpen
  \bibfield  {author} {\bibinfo {author} {\bibfnamefont {T.}~\bibnamefont {Matsumoto}}, \bibinfo {author} {\bibfnamefont {M.}~\bibnamefont {Kitazawa}},\ and\ \bibinfo {author} {\bibfnamefont {Y.}~\bibnamefont {Kohno}},\ }\bibfield  {title} {\bibinfo {title} {{Classifying topological charge in SU(3) Yang{\textendash}Mills theory with machine learning}},\ }\href {https://doi.org/10.1093/ptep/ptaa138} {\bibfield  {journal} {\bibinfo  {journal} {PTEP}\ }\textbf {\bibinfo {volume} {2021}},\ \bibinfo {pages} {023D01} (\bibinfo {year} {2021})},\ \Eprint {https://arxiv.org/abs/1909.06238} {arXiv:1909.06238 [hep-lat]} \BibitemShut {NoStop}%
\bibitem [{\citenamefont {Boyda}\ \emph {et~al.}(2021)\citenamefont {Boyda}, \citenamefont {Kanwar}, \citenamefont {Racani{\`e}re}, \citenamefont {Rezende}, \citenamefont {Albergo}, \citenamefont {Cranmer}, \citenamefont {Hackett},\ and\ \citenamefont {Shanahan}}]{Boyda:2020hsi}%
  \BibitemOpen
  \bibfield  {author} {\bibinfo {author} {\bibfnamefont {D.}~\bibnamefont {Boyda}}, \bibinfo {author} {\bibfnamefont {G.}~\bibnamefont {Kanwar}}, \bibinfo {author} {\bibfnamefont {S.}~\bibnamefont {Racani{\`e}re}}, \bibinfo {author} {\bibfnamefont {D.~J.}\ \bibnamefont {Rezende}}, \bibinfo {author} {\bibfnamefont {M.~S.}\ \bibnamefont {Albergo}}, \bibinfo {author} {\bibfnamefont {K.}~\bibnamefont {Cranmer}}, \bibinfo {author} {\bibfnamefont {D.~C.}\ \bibnamefont {Hackett}},\ and\ \bibinfo {author} {\bibfnamefont {P.~E.}\ \bibnamefont {Shanahan}},\ }\bibfield  {title} {\bibinfo {title} {{Sampling using $SU(N)$ gauge equivariant flows}},\ }\href {https://doi.org/10.1103/PhysRevD.103.074504} {\bibfield  {journal} {\bibinfo  {journal} {Phys. Rev. D}\ }\textbf {\bibinfo {volume} {103}},\ \bibinfo {pages} {074504} (\bibinfo {year} {2021})},\ \Eprint {https://arxiv.org/abs/2008.05456} {arXiv:2008.05456 [hep-lat]} \BibitemShut {NoStop}%
\bibitem [{\citenamefont {Abbott}\ \emph {et~al.}(2022)\citenamefont {Abbott} \emph {et~al.}}]{Abbott:2022zhs}%
  \BibitemOpen
  \bibfield  {author} {\bibinfo {author} {\bibfnamefont {R.}~\bibnamefont {Abbott}} \emph {et~al.},\ }\bibfield  {title} {\bibinfo {title} {{Gauge-equivariant flow models for sampling in lattice field theories with pseudofermions}},\ }\href {https://doi.org/10.1103/PhysRevD.106.074506} {\bibfield  {journal} {\bibinfo  {journal} {Phys. Rev. D}\ }\textbf {\bibinfo {volume} {106}},\ \bibinfo {pages} {074506} (\bibinfo {year} {2022})},\ \Eprint {https://arxiv.org/abs/2207.08945} {arXiv:2207.08945 [hep-lat]} \BibitemShut {NoStop}%
\bibitem [{\citenamefont {Favoni}\ \emph {et~al.}(2022)\citenamefont {Favoni}, \citenamefont {Ipp}, \citenamefont {M{\"u}ller},\ and\ \citenamefont {Schuh}}]{Favoni:2020reg}%
  \BibitemOpen
  \bibfield  {author} {\bibinfo {author} {\bibfnamefont {M.}~\bibnamefont {Favoni}}, \bibinfo {author} {\bibfnamefont {A.}~\bibnamefont {Ipp}}, \bibinfo {author} {\bibfnamefont {D.~I.}\ \bibnamefont {M{\"u}ller}},\ and\ \bibinfo {author} {\bibfnamefont {D.}~\bibnamefont {Schuh}},\ }\bibfield  {title} {\bibinfo {title} {{Lattice Gauge Equivariant Convolutional Neural Networks}},\ }\href {https://doi.org/10.1103/PhysRevLett.128.032003} {\bibfield  {journal} {\bibinfo  {journal} {Phys. Rev. Lett.}\ }\textbf {\bibinfo {volume} {128}},\ \bibinfo {pages} {032003} (\bibinfo {year} {2022})},\ \Eprint {https://arxiv.org/abs/2012.12901} {arXiv:2012.12901 [hep-lat]} \BibitemShut {NoStop}%
\bibitem [{\citenamefont {Nicoli}\ \emph {et~al.}(2021)\citenamefont {Nicoli}, \citenamefont {Anders}, \citenamefont {Funcke}, \citenamefont {Hartung}, \citenamefont {Jansen}, \citenamefont {Kessel}, \citenamefont {Nakajima},\ and\ \citenamefont {Stornati}}]{Nicoli:2020njz}%
  \BibitemOpen
  \bibfield  {author} {\bibinfo {author} {\bibfnamefont {K.~A.}\ \bibnamefont {Nicoli}}, \bibinfo {author} {\bibfnamefont {C.~J.}\ \bibnamefont {Anders}}, \bibinfo {author} {\bibfnamefont {L.}~\bibnamefont {Funcke}}, \bibinfo {author} {\bibfnamefont {T.}~\bibnamefont {Hartung}}, \bibinfo {author} {\bibfnamefont {K.}~\bibnamefont {Jansen}}, \bibinfo {author} {\bibfnamefont {P.}~\bibnamefont {Kessel}}, \bibinfo {author} {\bibfnamefont {S.}~\bibnamefont {Nakajima}},\ and\ \bibinfo {author} {\bibfnamefont {P.}~\bibnamefont {Stornati}},\ }\bibfield  {title} {\bibinfo {title} {{Estimation of Thermodynamic Observables in Lattice Field Theories with Deep Generative Models}},\ }\href {https://doi.org/10.1103/PhysRevLett.126.032001} {\bibfield  {journal} {\bibinfo  {journal} {Phys. Rev. Lett.}\ }\textbf {\bibinfo {volume} {126}},\ \bibinfo {pages} {032001} (\bibinfo {year} {2021})},\ \Eprint {https://arxiv.org/abs/2007.07115} {arXiv:2007.07115 [hep-lat]} \BibitemShut {NoStop}%
\bibitem [{\citenamefont {Wang}\ \emph {et~al.}(2024)\citenamefont {Wang}, \citenamefont {Aarts},\ and\ \citenamefont {Zhou}}]{Wang:2023exq}%
  \BibitemOpen
  \bibfield  {author} {\bibinfo {author} {\bibfnamefont {L.}~\bibnamefont {Wang}}, \bibinfo {author} {\bibfnamefont {G.}~\bibnamefont {Aarts}},\ and\ \bibinfo {author} {\bibfnamefont {K.}~\bibnamefont {Zhou}},\ }\bibfield  {title} {\bibinfo {title} {{Diffusion models as stochastic quantization in lattice field theory}},\ }\href {https://doi.org/10.1007/JHEP05(2024)060} {\bibfield  {journal} {\bibinfo  {journal} {JHEP}\ }\textbf {\bibinfo {volume} {05}},\ \bibinfo {pages} {060}},\ \Eprint {https://arxiv.org/abs/2309.17082} {arXiv:2309.17082 [hep-lat]} \BibitemShut {NoStop}%
\bibitem [{\citenamefont {Boyda}\ \emph {et~al.}(2022)\citenamefont {Boyda} \emph {et~al.}}]{Boyda:2022nmh}%
  \BibitemOpen
  \bibfield  {author} {\bibinfo {author} {\bibfnamefont {D.}~\bibnamefont {Boyda}} \emph {et~al.},\ }\bibfield  {title} {\bibinfo {title} {{Applications of Machine Learning to Lattice Quantum Field Theory}},\ }in\ \href@noop {} {\emph {\bibinfo {booktitle} {{Snowmass 2021}}}}\ (\bibinfo {year} {2022})\ \Eprint {https://arxiv.org/abs/2202.05838} {arXiv:2202.05838 [hep-lat]} \BibitemShut {NoStop}%
\bibitem [{\citenamefont {Albergo}\ \emph {et~al.}(2021)\citenamefont {Albergo}, \citenamefont {Boyda}, \citenamefont {Hackett}, \citenamefont {Kanwar}, \citenamefont {Cranmer}, \citenamefont {Racani{\`e}re}, \citenamefont {Rezende},\ and\ \citenamefont {Shanahan}}]{Albergo:2021vyo}%
  \BibitemOpen
  \bibfield  {author} {\bibinfo {author} {\bibfnamefont {M.~S.}\ \bibnamefont {Albergo}}, \bibinfo {author} {\bibfnamefont {D.}~\bibnamefont {Boyda}}, \bibinfo {author} {\bibfnamefont {D.~C.}\ \bibnamefont {Hackett}}, \bibinfo {author} {\bibfnamefont {G.}~\bibnamefont {Kanwar}}, \bibinfo {author} {\bibfnamefont {K.}~\bibnamefont {Cranmer}}, \bibinfo {author} {\bibfnamefont {S.}~\bibnamefont {Racani{\`e}re}}, \bibinfo {author} {\bibfnamefont {D.~J.}\ \bibnamefont {Rezende}},\ and\ \bibinfo {author} {\bibfnamefont {P.~E.}\ \bibnamefont {Shanahan}},\ }\href@noop {} {\bibinfo {title} {{Introduction to Normalizing Flows for Lattice Field Theory}}} (\bibinfo {year} {2021}),\ \Eprint {https://arxiv.org/abs/2101.08176} {arXiv:2101.08176 [hep-lat]} \BibitemShut {NoStop}%
\bibitem [{\citenamefont {Cranmer}\ \emph {et~al.}(2023)\citenamefont {Cranmer}, \citenamefont {Kanwar}, \citenamefont {Racani{\`e}re}, \citenamefont {Rezende},\ and\ \citenamefont {Shanahan}}]{Cranmer:2023xbe}%
  \BibitemOpen
  \bibfield  {author} {\bibinfo {author} {\bibfnamefont {K.}~\bibnamefont {Cranmer}}, \bibinfo {author} {\bibfnamefont {G.}~\bibnamefont {Kanwar}}, \bibinfo {author} {\bibfnamefont {S.}~\bibnamefont {Racani{\`e}re}}, \bibinfo {author} {\bibfnamefont {D.~J.}\ \bibnamefont {Rezende}},\ and\ \bibinfo {author} {\bibfnamefont {P.~E.}\ \bibnamefont {Shanahan}},\ }\bibfield  {title} {\bibinfo {title} {{Advances in machine-learning-based sampling motivated by lattice quantum chromodynamics}},\ }\href {https://doi.org/10.1038/s42254-023-00616-w} {\bibfield  {journal} {\bibinfo  {journal} {Nature Rev. Phys.}\ }\textbf {\bibinfo {volume} {5}},\ \bibinfo {pages} {526} (\bibinfo {year} {2023})},\ \Eprint {https://arxiv.org/abs/2309.01156} {arXiv:2309.01156 [hep-lat]} \BibitemShut {NoStop}%
\bibitem [{\citenamefont {Tomiya}(2025)}]{Tomiya:2025quf}%
  \BibitemOpen
  \bibfield  {author} {\bibinfo {author} {\bibfnamefont {A.}~\bibnamefont {Tomiya}},\ }\bibfield  {title} {\bibinfo {title} {{Machine Learning for Lattice QCD}},\ }\href {https://doi.org/10.7566/JPSJ.94.031006} {\bibfield  {journal} {\bibinfo  {journal} {J. Phys. Soc. Jap.}\ }\textbf {\bibinfo {volume} {94}},\ \bibinfo {pages} {031006} (\bibinfo {year} {2025})}\BibitemShut {NoStop}%
\bibitem [{\citenamefont {Yasunaga}\ \emph {et~al.}(2026)\citenamefont {Yasunaga}, \citenamefont {Yoshimura}, \citenamefont {Tomiya},\ and\ \citenamefont {Nagai}}]{Yasunaga:2026smj}%
  \BibitemOpen
  \bibfield  {author} {\bibinfo {author} {\bibfnamefont {S.}~\bibnamefont {Yasunaga}}, \bibinfo {author} {\bibfnamefont {K.}~\bibnamefont {Yoshimura}}, \bibinfo {author} {\bibfnamefont {A.}~\bibnamefont {Tomiya}},\ and\ \bibinfo {author} {\bibfnamefont {Y.}~\bibnamefont {Nagai}},\ }\bibfield  {title} {\bibinfo {title} {{Parameter Optimization of Domain-Wall Fermion using Machine Learning}},\ }in\ \href@noop {} {\emph {\bibinfo {booktitle} {{42th International Symposium on Lattice Field Theory}}}}\ (\bibinfo {year} {2026})\ \Eprint {https://arxiv.org/abs/2603.16329} {arXiv:2603.16329 [hep-lat]} \BibitemShut {NoStop}%
\bibitem [{\citenamefont {Misumi}(2026)}]{Misumi:2026zbe}%
  \BibitemOpen
  \bibfield  {author} {\bibinfo {author} {\bibfnamefont {T.}~\bibnamefont {Misumi}},\ }\href@noop {} {\bibinfo {title} {{Lattice fermion formulation via Physics-Informed Neural Networks: Ginsparg-Wilson relation and Overlap fermions}}} (\bibinfo {year} {2026}),\ \Eprint {https://arxiv.org/abs/2605.06022} {arXiv:2605.06022 [hep-lat]} \BibitemShut {NoStop}%
\bibitem [{\citenamefont {Wilson}(1974)}]{Wilson:1974sk}%
  \BibitemOpen
  \bibfield  {author} {\bibinfo {author} {\bibfnamefont {K.~G.}\ \bibnamefont {Wilson}},\ }\bibfield  {title} {\bibinfo {title} {{Confinement of Quarks}},\ }\href {https://doi.org/10.1103/PhysRevD.10.2445} {\bibfield  {journal} {\bibinfo  {journal} {Phys. Rev. D}\ }\textbf {\bibinfo {volume} {10}},\ \bibinfo {pages} {2445} (\bibinfo {year} {1974})}\BibitemShut {NoStop}%
\bibitem [{\citenamefont {Munster}\ and\ \citenamefont {Weisz}(1980)}]{Munster:1980vk}%
  \BibitemOpen
  \bibfield  {author} {\bibinfo {author} {\bibfnamefont {G.}~\bibnamefont {Munster}}\ and\ \bibinfo {author} {\bibfnamefont {P.}~\bibnamefont {Weisz}},\ }\bibfield  {title} {\bibinfo {title} {{Estimate of the Relation Between Scale Parameters and the String Tension by Strong Coupling Methods}},\ }\href {https://doi.org/10.1016/0370-2693(80)90225-7} {\bibfield  {journal} {\bibinfo  {journal} {Phys. Lett. B}\ }\textbf {\bibinfo {volume} {96}},\ \bibinfo {pages} {119} (\bibinfo {year} {1980})},\ \bibinfo {note} {[Erratum: Phys.Lett.B 100, 519 (1981)]}\BibitemShut {NoStop}%
\bibitem [{\citenamefont {Drouffe}\ and\ \citenamefont {Zuber}(1981)}]{Drouffe:1980dp}%
  \BibitemOpen
  \bibfield  {author} {\bibinfo {author} {\bibfnamefont {J.~M.}\ \bibnamefont {Drouffe}}\ and\ \bibinfo {author} {\bibfnamefont {J.~B.}\ \bibnamefont {Zuber}},\ }\bibfield  {title} {\bibinfo {title} {{Roughening Transition in Lattice Gauge Theories in Arbitrary Dimension. 2. The Groups $Z$(3), U(1), SU(2), SU(3)}},\ }\href {https://doi.org/10.1016/0550-3213(81)90419-3} {\bibfield  {journal} {\bibinfo  {journal} {Nucl. Phys. B}\ }\textbf {\bibinfo {volume} {180}},\ \bibinfo {pages} {264} (\bibinfo {year} {1981})}\BibitemShut {NoStop}%
\bibitem [{\citenamefont {Munster}(1981)}]{Munster:1981es}%
  \BibitemOpen
  \bibfield  {author} {\bibinfo {author} {\bibfnamefont {G.}~\bibnamefont {Munster}},\ }\bibfield  {title} {\bibinfo {title} {{Strong Coupling Expansions for the Mass Gap in Lattice Gauge Theories}},\ }\href {https://doi.org/10.1016/0550-3213(81)90570-8} {\bibfield  {journal} {\bibinfo  {journal} {Nucl. Phys. B}\ }\textbf {\bibinfo {volume} {190}},\ \bibinfo {pages} {439} (\bibinfo {year} {1981})},\ \bibinfo {note} {[Addendum: Nucl.Phys.B 200, 536--538 (1982), Erratum: Nucl.Phys.B 205, 648--648 (1982)]}\BibitemShut {NoStop}%
\bibitem [{\citenamefont {Creutz}(2023)}]{Creutz_2023}%
  \BibitemOpen
  \bibfield  {author} {\bibinfo {author} {\bibfnamefont {M.}~\bibnamefont {Creutz}},\ }\href@noop {} {\emph {\bibinfo {title} {Quarks, Gluons and Lattices}}},\ Cambridge Monographs on Mathematical Physics\ (\bibinfo  {publisher} {Cambridge University Press},\ \bibinfo {year} {2023})\BibitemShut {NoStop}%
\bibitem [{\citenamefont {Rothe}(1992)}]{Rothe:1992nt}%
  \BibitemOpen
  \bibfield  {author} {\bibinfo {author} {\bibfnamefont {H.~J.}\ \bibnamefont {Rothe}},\ }\href@noop {} {\emph {\bibinfo {title} {{Lattice gauge theories: An Introduction}}}},\ Vol.~\bibinfo {volume} {43}\ (\bibinfo  {publisher} {World Scientific},\ \bibinfo {year} {1992})\BibitemShut {NoStop}%
\bibitem [{\citenamefont {G{\"u}lpers}\ \emph {et~al.}(2014)\citenamefont {G{\"u}lpers}, \citenamefont {von Hippel},\ and\ \citenamefont {Wittig}}]{Gulpers:2013uca}%
  \BibitemOpen
  \bibfield  {author} {\bibinfo {author} {\bibfnamefont {V.}~\bibnamefont {G{\"u}lpers}}, \bibinfo {author} {\bibfnamefont {G.}~\bibnamefont {von Hippel}},\ and\ \bibinfo {author} {\bibfnamefont {H.}~\bibnamefont {Wittig}},\ }\bibfield  {title} {\bibinfo {title} {{Scalar pion form factor in two-flavor lattice QCD}},\ }\href {https://doi.org/10.1103/PhysRevD.89.094503} {\bibfield  {journal} {\bibinfo  {journal} {Phys. Rev. D}\ }\textbf {\bibinfo {volume} {89}},\ \bibinfo {pages} {094503} (\bibinfo {year} {2014})},\ \Eprint {https://arxiv.org/abs/1309.2104} {arXiv:1309.2104 [hep-lat]} \BibitemShut {NoStop}%
\bibitem [{\citenamefont {Zhou}\ \emph {et~al.}(2018)\citenamefont {Zhou}, \citenamefont {Cheng}, \citenamefont {Xiong},\ and\ \citenamefont {Zhang}}]{Zhou:2018}%
  \BibitemOpen
  \bibfield  {author} {\bibinfo {author} {\bibfnamefont {J.-L.}\ \bibnamefont {Zhou}}, \bibinfo {author} {\bibfnamefont {Z.}~\bibnamefont {Cheng}}, \bibinfo {author} {\bibfnamefont {G.-Y.}\ \bibnamefont {Xiong}},\ and\ \bibinfo {author} {\bibfnamefont {J.-B.}\ \bibnamefont {Zhang}},\ }\bibfield  {title} {\bibinfo {title} {Hopping parameter expansion technique in noise method for disconnected quark loops},\ }\href {https://doi.org/10.1088/0256-307X/35/4/041101} {\bibfield  {journal} {\bibinfo  {journal} {Chin. Phys. Lett.}\ }\textbf {\bibinfo {volume} {35}},\ \bibinfo {pages} {041101} (\bibinfo {year} {2018})}\BibitemShut {NoStop}%
\bibitem [{\citenamefont {Hasenbusch}(2018)}]{Hasenbusch:2018iuw}%
  \BibitemOpen
  \bibfield  {author} {\bibinfo {author} {\bibfnamefont {M.}~\bibnamefont {Hasenbusch}},\ }\bibfield  {title} {\bibinfo {title} {{Exploiting the hopping parameter expansion in the hybrid Monte Carlo simulation of lattice QCD with two degenerate flavors of Wilson fermions}},\ }\href {https://doi.org/10.1103/PhysRevD.97.114512} {\bibfield  {journal} {\bibinfo  {journal} {Phys. Rev. D}\ }\textbf {\bibinfo {volume} {97}},\ \bibinfo {pages} {114512} (\bibinfo {year} {2018})},\ \Eprint {https://arxiv.org/abs/1805.03560} {arXiv:1805.03560 [hep-lat]} \BibitemShut {NoStop}%
\bibitem [{\citenamefont {Baral}\ \emph {et~al.}(2019)\citenamefont {Baral}, \citenamefont {Whyte}, \citenamefont {Wilcox},\ and\ \citenamefont {Morgan}}]{Baral:2019mkt}%
  \BibitemOpen
  \bibfield  {author} {\bibinfo {author} {\bibfnamefont {S.}~\bibnamefont {Baral}}, \bibinfo {author} {\bibfnamefont {T.}~\bibnamefont {Whyte}}, \bibinfo {author} {\bibfnamefont {W.}~\bibnamefont {Wilcox}},\ and\ \bibinfo {author} {\bibfnamefont {R.~B.}\ \bibnamefont {Morgan}},\ }\bibfield  {title} {\bibinfo {title} {{Disconnected Loop Subtraction Methods in Lattice QCD}},\ }\href {https://doi.org/10.1016/j.cpc.2019.03.011} {\bibfield  {journal} {\bibinfo  {journal} {Comput. Phys. Commun.}\ }\textbf {\bibinfo {volume} {241}},\ \bibinfo {pages} {64} (\bibinfo {year} {2019})},\ \Eprint {https://arxiv.org/abs/1903.06986} {arXiv:1903.06986 [hep-lat]} \BibitemShut {NoStop}%
\bibitem [{\citenamefont {Giusti}\ \emph {et~al.}(2019)\citenamefont {Giusti}, \citenamefont {Harris}, \citenamefont {Nada},\ and\ \citenamefont {Schaefer}}]{Giusti:2019kff}%
  \BibitemOpen
  \bibfield  {author} {\bibinfo {author} {\bibfnamefont {L.}~\bibnamefont {Giusti}}, \bibinfo {author} {\bibfnamefont {T.}~\bibnamefont {Harris}}, \bibinfo {author} {\bibfnamefont {A.}~\bibnamefont {Nada}},\ and\ \bibinfo {author} {\bibfnamefont {S.}~\bibnamefont {Schaefer}},\ }\bibfield  {title} {\bibinfo {title} {{Frequency-splitting estimators of single-propagator traces}},\ }\href {https://doi.org/10.1140/epjc/s10052-019-7049-0} {\bibfield  {journal} {\bibinfo  {journal} {Eur. Phys. J. C}\ }\textbf {\bibinfo {volume} {79}},\ \bibinfo {pages} {586} (\bibinfo {year} {2019})},\ \Eprint {https://arxiv.org/abs/1903.10447} {arXiv:1903.10447 [hep-lat]} \BibitemShut {NoStop}%
\bibitem [{\citenamefont {Djukanovic}\ \emph {et~al.}(2024)\citenamefont {Djukanovic}, \citenamefont {von Hippel}, \citenamefont {Meyer}, \citenamefont {Ottnad}, \citenamefont {Salg},\ and\ \citenamefont {Wittig}}]{Djukanovic:2023beb}%
  \BibitemOpen
  \bibfield  {author} {\bibinfo {author} {\bibfnamefont {D.}~\bibnamefont {Djukanovic}}, \bibinfo {author} {\bibfnamefont {G.}~\bibnamefont {von Hippel}}, \bibinfo {author} {\bibfnamefont {H.~B.}\ \bibnamefont {Meyer}}, \bibinfo {author} {\bibfnamefont {K.}~\bibnamefont {Ottnad}}, \bibinfo {author} {\bibfnamefont {M.}~\bibnamefont {Salg}},\ and\ \bibinfo {author} {\bibfnamefont {H.}~\bibnamefont {Wittig}},\ }\bibfield  {title} {\bibinfo {title} {{Electromagnetic form factors of the nucleon from Nf=2+1 lattice QCD}},\ }\href {https://doi.org/10.1103/PhysRevD.109.094510} {\bibfield  {journal} {\bibinfo  {journal} {Phys. Rev. D}\ }\textbf {\bibinfo {volume} {109}},\ \bibinfo {pages} {094510} (\bibinfo {year} {2024})},\ \Eprint {https://arxiv.org/abs/2309.06590} {arXiv:2309.06590 [hep-lat]} \BibitemShut {NoStop}%
\bibitem [{\citenamefont {Philipsen}\ and\ \citenamefont {Pinke}(2014)}]{Philipsen:2014rpa}%
  \BibitemOpen
  \bibfield  {author} {\bibinfo {author} {\bibfnamefont {O.}~\bibnamefont {Philipsen}}\ and\ \bibinfo {author} {\bibfnamefont {C.}~\bibnamefont {Pinke}},\ }\bibfield  {title} {\bibinfo {title} {{Nature of the Roberge-Weiss transition in $N_f=2$ QCD with Wilson fermions}},\ }\href {https://doi.org/10.1103/PhysRevD.89.094504} {\bibfield  {journal} {\bibinfo  {journal} {Phys. Rev. D}\ }\textbf {\bibinfo {volume} {89}},\ \bibinfo {pages} {094504} (\bibinfo {year} {2014})},\ \Eprint {https://arxiv.org/abs/1402.0838} {arXiv:1402.0838 [hep-lat]} \BibitemShut {NoStop}%
\bibitem [{\citenamefont {Cuteri}\ \emph {et~al.}(2021)\citenamefont {Cuteri}, \citenamefont {Philipsen}, \citenamefont {Sch\"on},\ and\ \citenamefont {Sciarra}}]{Cuteri:2020yke}%
  \BibitemOpen
  \bibfield  {author} {\bibinfo {author} {\bibfnamefont {F.}~\bibnamefont {Cuteri}}, \bibinfo {author} {\bibfnamefont {O.}~\bibnamefont {Philipsen}}, \bibinfo {author} {\bibfnamefont {A.}~\bibnamefont {Sch\"on}},\ and\ \bibinfo {author} {\bibfnamefont {A.}~\bibnamefont {Sciarra}},\ }\bibfield  {title} {\bibinfo {title} {{Deconfinement critical point of lattice QCD with $N_f$=2 Wilson fermions}},\ }\href {https://doi.org/10.1103/PhysRevD.103.014513} {\bibfield  {journal} {\bibinfo  {journal} {Phys. Rev. D}\ }\textbf {\bibinfo {volume} {103}},\ \bibinfo {pages} {014513} (\bibinfo {year} {2021})},\ \Eprint {https://arxiv.org/abs/2009.14033} {arXiv:2009.14033 [hep-lat]} \BibitemShut {NoStop}%
\bibitem [{\citenamefont {Philipsen}(2021)}]{Philipsen:2021qji}%
  \BibitemOpen
  \bibfield  {author} {\bibinfo {author} {\bibfnamefont {O.}~\bibnamefont {Philipsen}},\ }\bibfield  {title} {\bibinfo {title} {{Lattice Constraints on the QCD Chiral Phase Transition at Finite Temperature and Baryon Density}},\ }\href {https://doi.org/10.3390/sym13112079} {\bibfield  {journal} {\bibinfo  {journal} {Symmetry}\ }\textbf {\bibinfo {volume} {13}},\ \bibinfo {pages} {2079} (\bibinfo {year} {2021})},\ \Eprint {https://arxiv.org/abs/2111.03590} {arXiv:2111.03590 [hep-lat]} \BibitemShut {NoStop}%
\bibitem [{\citenamefont {Kiyohara}\ \emph {et~al.}(2021)\citenamefont {Kiyohara}, \citenamefont {Kitazawa}, \citenamefont {Ejiri},\ and\ \citenamefont {Kanaya}}]{Kiyohara:2021smr}%
  \BibitemOpen
  \bibfield  {author} {\bibinfo {author} {\bibfnamefont {A.}~\bibnamefont {Kiyohara}}, \bibinfo {author} {\bibfnamefont {M.}~\bibnamefont {Kitazawa}}, \bibinfo {author} {\bibfnamefont {S.}~\bibnamefont {Ejiri}},\ and\ \bibinfo {author} {\bibfnamefont {K.}~\bibnamefont {Kanaya}},\ }\bibfield  {title} {\bibinfo {title} {{Finite-size scaling around the critical point in the heavy quark region of QCD}},\ }\href {https://doi.org/10.1103/PhysRevD.104.114509} {\bibfield  {journal} {\bibinfo  {journal} {Phys. Rev. D}\ }\textbf {\bibinfo {volume} {104}},\ \bibinfo {pages} {114509} (\bibinfo {year} {2021})},\ \Eprint {https://arxiv.org/abs/2108.00118} {arXiv:2108.00118 [hep-lat]} \BibitemShut {NoStop}%
\bibitem [{\citenamefont {Ashikawa}\ \emph {et~al.}(2024)\citenamefont {Ashikawa}, \citenamefont {Kitazawa}, \citenamefont {Ejiri},\ and\ \citenamefont {Kanaya}}]{Ashikawa:2024njc}%
  \BibitemOpen
  \bibfield  {author} {\bibinfo {author} {\bibfnamefont {R.}~\bibnamefont {Ashikawa}}, \bibinfo {author} {\bibfnamefont {M.}~\bibnamefont {Kitazawa}}, \bibinfo {author} {\bibfnamefont {S.}~\bibnamefont {Ejiri}},\ and\ \bibinfo {author} {\bibfnamefont {K.}~\bibnamefont {Kanaya}},\ }\bibfield  {title} {\bibinfo {title} {{High-precision analysis of the critical point in heavy-quark QCD at Nt=6}},\ }\href {https://doi.org/10.1103/PhysRevD.110.074508} {\bibfield  {journal} {\bibinfo  {journal} {Phys. Rev. D}\ }\textbf {\bibinfo {volume} {110}},\ \bibinfo {pages} {074508} (\bibinfo {year} {2024})},\ \Eprint {https://arxiv.org/abs/2407.09156} {arXiv:2407.09156 [hep-lat]} \BibitemShut {NoStop}%
\bibitem [{\citenamefont {Fredkin}(1960)}]{Fredkin1960}%
  \BibitemOpen
  \bibfield  {author} {\bibinfo {author} {\bibfnamefont {E.}~\bibnamefont {Fredkin}},\ }\bibfield  {title} {\bibinfo {title} {Trie memory},\ }\href {https://doi.org/10.1145/367390.367400} {\bibfield  {journal} {\bibinfo  {journal} {Communications of the ACM}\ }\textbf {\bibinfo {volume} {3}},\ \bibinfo {pages} {490} (\bibinfo {year} {1960})}\BibitemShut {NoStop}%
\bibitem [{Wad()}]{WadaHPECode}%
  \BibitemOpen
  \href@noop {} {}\bibinfo {howpublished} {\url{https://github.com/WTatsuya2000/HO-HPE}}\BibitemShut {NoStop}%
\bibitem [{\citenamefont {Gattringer}\ and\ \citenamefont {Lang}(2010)}]{Gattringer:2010zz}%
  \BibitemOpen
  \bibfield  {author} {\bibinfo {author} {\bibfnamefont {C.}~\bibnamefont {Gattringer}}\ and\ \bibinfo {author} {\bibfnamefont {C.~B.}\ \bibnamefont {Lang}},\ }\href {https://doi.org/10.1007/978-3-642-01850-3} {\emph {\bibinfo {title} {{Quantum chromodynamics on the lattice}}}},\ Vol.\ \bibinfo {volume} {788}\ (\bibinfo  {publisher} {Springer},\ \bibinfo {address} {Berlin},\ \bibinfo {year} {2010})\BibitemShut {NoStop}%
\bibitem [{\citenamefont {Wakabayashi}\ \emph {et~al.}(2022)\citenamefont {Wakabayashi}, \citenamefont {Ejiri}, \citenamefont {Kanaya},\ and\ \citenamefont {Kitazawa}}]{Wakabayashi:2021eye}%
  \BibitemOpen
  \bibfield  {author} {\bibinfo {author} {\bibfnamefont {N.}~\bibnamefont {Wakabayashi}}, \bibinfo {author} {\bibfnamefont {S.}~\bibnamefont {Ejiri}}, \bibinfo {author} {\bibfnamefont {K.}~\bibnamefont {Kanaya}},\ and\ \bibinfo {author} {\bibfnamefont {M.}~\bibnamefont {Kitazawa}},\ }\bibfield  {title} {\bibinfo {title} {{Scope and convergence of the hopping parameter expansion in finite-temperature quantum chromodynamics with heavy quarks around the critical point}},\ }\href {https://doi.org/10.1093/ptep/ptac019} {\bibfield  {journal} {\bibinfo  {journal} {PTEP}\ }\textbf {\bibinfo {volume} {2022}},\ \bibinfo {pages} {033B05} (\bibinfo {year} {2022})},\ \Eprint {https://arxiv.org/abs/2112.06340} {arXiv:2112.06340 [hep-lat]} \BibitemShut {NoStop}%
\bibitem [{\citenamefont {Burnside}(1897)}]{Burnside:1897}%
  \BibitemOpen
  \bibfield  {author} {\bibinfo {author} {\bibfnamefont {W.}~\bibnamefont {Burnside}},\ }\href@noop {} {\emph {\bibinfo {title} {{Theory of Groups of Finite Order}}}}\ (\bibinfo  {publisher} {Cambridge University Press},\ \bibinfo {address} {Cambridge},\ \bibinfo {year} {1897})\ \bibinfo {note} {2nd ed. 1911; Dover reprint 1955}\BibitemShut {NoStop}%
\end{thebibliography}%
\clearpage
\renewcommand{\arraystretch}{1.08}

\begin{center}
\textbf{Supplemental Material:\\
Tables for the shape classification for $W(4)$, $W(6)$, and $W(8)$}
\end{center}

\begin{table}[h]
\caption{Shape classification of $W(4)$.}
\label{tab:W4sm}
\begin{ruledtabular}
\begin{tabular}{rlrrrl}
$j$ & trajectory & $M_j$ & $S_j$ & $D_j$ & type \\
\hline
$1$ & $12\bar{1}\bar{2}$ & $6$ & $1$ & $-8$ & plaquette

\end{tabular}
\end{ruledtabular}
\end{table}

\begin{table}[h]
\caption{Shape classification of $W(6)$.}
\label{tab:W6sm}
\begin{ruledtabular}
\begin{tabular}{rlrrrl}
$j$ & trajectory & $M_j$ & $S_j$ & $D_j$ & type \\
\hline
$1$ & $122\bar{1}\bar{2}\bar{2}$ & $12$ & $1$ & $-32$ & rectangle \\
$2$ & $123\bar{1}\bar{2}\bar{3}$ & $48$ & $1$ & $-16$ & chair \\
$3$ & $123\bar{1}\bar{3}\bar{2}$ & $16$ & $1$ & $-16$ & crown 

\end{tabular}
\end{ruledtabular}
\end{table}

\begin{table}[h]
\caption{Shape classification of $W(8)$.}
\label{tab:W8sm}
\begin{ruledtabular}
\begin{tabular}{rlrrrl}
$j$ & trajectory & $M_j$ & $S_j$ & $D_j$ & type \\
\hline
$1$ & $1112\bar1\bar1\bar1\bar2$ & $12$ & $1$ & $-128$ & U=U \\
$2$ & $12\bar1\bar2\bar121\bar2$ & $12$ & $1$ & $32$ & PP \\
$3$ & $12\bar1\bar2\bar1\bar212$ & $12$ & $1$ & $-32$ & PP \\
$4$ & $12\bar1\bar2\bar2\bar121$ & $12$ & $1$ & $64$ & PP \\
$5$ & $12\bar1\bar212\bar1\bar2$ & $6$ & $2$ & $32$ & ${\rm P}^2$ \\
$6$ & $112\bar12\bar1\bar2\bar2$ & $24$ & $1$ & $-64$ & ULU \\
$7$ & $1122\bar1\bar1\bar2\bar2$ & $6$ & $1$ & $-128$ & $\Box$ \\
\hline
$8$ & $12\bar1\bar232\bar3\bar2$ & $48$ & $1$ & $32$ & PP \\
$9$ & $12\bar1\bar23\bar1\bar31$ & $96$ & $1$ & $32$ & PP \\
$10$ & $12\bar1\bar2\bar131\bar3$ & $96$ & $1$ & $0$ & PP \\
$11$ & $112\bar1\bar13\bar2\bar3$ & $96$ & $1$ & $-64$ & U=U \\
$12$ & $123\bar2\bar12\bar3\bar2$ & $24$ & $1$ & $-32$ & U=U \\
$13$ & $123\bar2\bar1\bar2\bar32$ & $48$ & $1$ & $-32$ & U=U \\
$14$ & $112\bar13\bar1\bar3\bar2$ & $192$ & $1$ & $-32$ & ULU \\
$15$ & $123\bar2\bar2\bar12\bar3$ & $96$ & $1$ & $-32$ & ULU \\
$16$ & $1123\bar1\bar1\bar3\bar2$ & $48$ & $1$ & $-64$ & L==L \\
$17$ & $1123\bar1\bar1\bar2\bar3$ & $48$ & $1$ & $-64$ & L==L \\
$18$ & $1233\bar1\bar3\bar2\bar3$ & $192$ & $1$ & $-32$ & crown + U \\
\hline
$19$ & $12\bar1\bar234\bar3\bar4$ & $96$ & $1$ & $16$ & PP \\
$20$ & $123\bar2\bar14\bar3\bar4$ & $96$ & $1$ & $-32$ & U=U \\
$21$ & $123\bar24\bar1\bar4\bar3$ & $192$ & $1$ & $-16$ & ULU \\
$22$ & $123\bar14\bar3\bar2\bar4$ & $48$ & $1$ & $-32$ & $\Sigma\Sigma$ \\
$23$ & $123\bar14\bar2\bar3\bar4$ & $96$ & $1$ & $-32$ & $\Sigma\Sigma$ \\
$24$ & $1234\bar1\bar4\bar2\bar3$ & $192$ & $1$ & $-16$ & crown + U \\
$25$ & $1234\bar1\bar2\bar3\bar4$ & $24$ & $1$ & $-16$ & 4d-crown
\end{tabular}
\end{ruledtabular}
\end{table}

\end{document}